\documentclass[12pt,preprint]{aastex}

\def\be{\begin{equation}}
\def\ee{\end{equation}}

\def\ergs{{\rm\,erg\,s^{-1}}}
\def\msun{M_{\odot}}

\def\ergs{\rm \,erg\,s^{-1}}
\def\be{\begin{equation}}
\def\ee{\end{equation}}
\catcode`\@=11 
\def\@versim#1#2{\vcenter{\offinterlineskip
        \ialign{$\m@th#1\hfil##\hfil$\crcr#2\crcr\sim\crcr } }}
\def\lsim{\mathrel{\mathpalette\@versim<}}
\def\gsim{\mathrel{\mathpalette\@versim>}}

\shorttitle{Radio--X-ray Correlation and Quiescent State}
\shortauthors{Feng Yuan and Wei Cui}
\begin{document}

\title{Radio---X-ray Correlation and the ``Quiescent State'' of Black Hole
Sources}

\author{Feng Yuan and Wei Cui}
\affil{Department of Physics, Purdue University, West Lafayette, IN 47907}
\email{fyuan@physics.purdue.edu; cui@physics.purdue.edu}

\begin{abstract}

Recently a correlation between the radio and X-ray luminosities is found,
$L_{\rm R}\propto L_{\rm X}^{0.7}$, in black hole sources
including black hole candidates in our galaxy and active galactic nuclei.
We first show that the correlation can be understood
in the context of an accretion-jet model developed for explaining
the spectral and timing properties of XTE J1118+480. 
More importantly, we show that when the X-ray luminosity is 
below a critical value, $\la (10^{-5}$--$10^{-6}) L_{\rm Edd}$, 
if the jet persists, the correlation should 
turn and become steeper, $L_{\rm R}\propto 
L_{\rm X}^{1.23}$, and the X-ray radiation of the system should be dominated
by the emission from the jet, rather than by the accretion flow. 
Possible observational evidence for our predictions is presented
and future observations to further test our predictions are proposed.

\end{abstract}


\section{Introduction}

In the low/hard state, the radio spectrum of black hole candidates (BHCs)
is usually flat or even inverted, which is often taken as evidence
for the presence of jets (Fender 2004 and references therein).
This is because, on one hand, such a spectrum is characteristic
of jets in active galactic nuclei (AGNs; e.g., Blandford \& 
K\"onigl 1979); on the other hand, it is difficult to explain it
by invoking emission from the underlying accretion flow (Yuan, Cui,
\& Narayan 2005, hereafter YCN05). A
strong correlation between the radio and X-ray luminosities of 
BHCs in the hard state has been found recently
(Corbel et al. 2003; Gallo, Fender, \& Pooley 2003),
$L_{\rm R} \propto L_{\rm X}^{0.7}$, where $L_{\rm R}$
is the radio luminosity at 8.6 GHz and $L_{\rm X}$ is the 2-11 keV X-ray 
luminosity. The correlation extends more than three
decades in $L_{\rm X}$, from $L_{\rm X} \ga 10^{-2} L_{\rm Edd}$ to
$L_{\rm X} \la 10^{-5}L_{\rm Edd}$ ($L_{\rm Edd}$ is the 
Eddington luminosity). The lowest luminosity is close
to the quiescent state luminosity of V404 Cyg, but is
still much higher than that of typical
BHCs (e.g., Kong et al. 2002; McClintock et al. 2003).
The correlation also holds for AGNs
(Merloni, Heinz, \& Di Matteo 2003, hereafter MHD03; 
Falcke, K\"ording, \& Markoff 2004).

It is generally thought that the X-ray emission from BHCs originates in the
accretion flows (see a review by Narayan 2004). 
The observed radio---X-ray correlation strongly implies a casual 
connection between the accretion flow and jet.
Recently coupled accretion-jet models have been proposed
and applied to the hard state of XTE J1118+480, 
a source to which we
have almost the best spectral and timing observational results among all BHCs
(YCN05; Malzac, Merloni, \& Fabian 2004).
In these models, the accretion flow is composed of an inner
ADAF-like hot accretion flow and an outer standard
thin disk (Esin, McClintock \& Narayan 1997; Malzac, Merloni, 
\& Fabian also discuss the possibility that the X-rays may come from a
patchy corona rather than an ADAF). An additional jet component is 
assumed to form at the innermost region of the accretion flow.
The X-ray emission is dominated by the hot accretion 
flow, while the radio emission comes
from the jet. The accretion-jet models can not only
explain the broadband spectral energy distribution of XTE J1118+480 ranging
from radio to X-ray (YCN05), but also account for most of its 
complicated timing features (YCN05; Malzac, Merloni, \& Fabian 2004).

It is natural to ask whether the models can explain 
the observed radio---X-ray correlation in a quantitative manner, or 
what constraints the correlation can put on the
models. One critical parameter in the problem 
is how the fraction of the accreted matter that goes into the jet, 
$\eta~(\equiv \dot{M}_{\rm jet}/\dot{M})$, changes with $\dot{M}$. 
Unfortunately, no good theoretical constraint can be put on it
due to our poor understanding of the jet formation. Assuming 
$\eta$ is constant, MHD03 find that an ADAF-jet model is 
roughly consistent with the observations at the 3$\sigma$ level. 

In this paper we begin by investigating this 
question again. Instead of assuming a constant  $\eta$, we 
investigate what functional form of $\eta(\dot{M})$ is required 
to explain the radio---X-ray correlation (\S2.1). 
Such a study may supply us with some clue on the jet formation mechanism.
In \S2.2 and \S2.3, we investigate 
what the correlation will be below the lowest observed
X-ray luminosity ($\sim 10^{-5}L_{\rm Edd}$). 
We find that when $L_{\rm X} \la 
10^{-5}$--$10^{-6}L_{\rm Edd}$, the radio---X-ray correlation
will become much steeper, $L_{\rm R}\propto L_{\rm X}^{1.23}$, and 
the X-ray emission of the source should be dominated
by the jet, rather than by the accretion flow.

\section{Model}

\subsection{Radio---X-ray correlation in the context of accretion-jet model}

There are some uncertainties in the normalization
of the observed radio---X-ray correlation (e.g., Gallo, Fender \& Pooley 2003).
Without losing generality, in our calculation we determine the normalization
from the observed outburst-state radio and X-ray fluxes of 
XTE J1118+480 (see YCN05 for the observational
data). Then the observed correlation is expressed as,
\be
\left(\frac{L_{\rm R}}{L_{\rm Edd}}\right) = 10^{-7.34}\left(\frac
{L_{\rm X}}{L_{\rm Edd}}\right)^{0.7}.
\ee
This is shown by the segment ``AB'' in Figure 1. 
The point ``A'' corresponds
to the outburst state of XTE J1118+480, and the point ``B'' 
the lowest X-ray luminosity to which the observed
correlation in BHCs extends, $\sim 10^{-5.2}L_{\rm Edd}$
(Gallo, Fender \& Pooley 2003).

The details of the accretion-jet model are described in YCN05.
Briefly, the accretion flow is described as a geometrically thin cool
disk outside a transition radius and a geometrically-thick hot accretion flow  
(i.e, ADAF) inside the transition radius. 
The effect of outflow/convection
is taken into account in calculating the dynamics of the hot accretion flow.
The main parameters are $\alpha=0.3, \beta=0.9$, and $\delta=0.5$.
Near the black hole, we assume that a fraction of the accretion flow,
$\eta$, is transferred into the vertical direction to form a jet. 
The half-opening angle of the jet is $\phi=0.1$ and 
the bulk Lorentz factor is $\Gamma_{\rm j}=1.2$.
Within the jet, internal shocks occur due to the collision of 
shells with different velocities. These shocks accelerate a 
fraction of the electrons into a power-law energy distribution. 
The steady state energy distribution of the accelerated
electrons is self-consistently determined, taking into account the effect 
of radiative cooling. Following the widely adopted approach
in the study of GRBs, the energy density of accelerated electrons and
amplified magnetic field is determined by two parameters,
$\epsilon_e$ and $\epsilon_B$, which describe 
the fraction of the shock energy going into electrons and magnetic 
field, respectively. The values of $\epsilon_e$ and $\epsilon_B$ are
0.06 and 0.02, respectively, which are well within the range of typical range 
obtained in GRB afterglows (see YCN05 for details). We then calculate the 
synchrotron emission from these 
accelerated electrons. Of course, like any other jet models 
published in the literature, our jet model is only phenomenological 
because the physics of jet formation is still poorly understood.

We calculate the values of $L_{\rm R}$ and $L_{\rm X}$ at various
$\dot{M}$, adjusting $\eta$ so that eq. (1) can be satisfied. 
The radio luminosity $L_{\rm R}$ is always dominated by the 
emission from the jet (by optically-thick synchrotron emission)
while $L_{\rm X}$ is the sum of the
emissions from the accretion flow (by thermal Comptonization 
and bremsstrahlung emissions) and jet (by
optically-thin synchrotron emission).
We assume that the intrinsic physics of accretion and jet 
does not depends on $\dot{M}$, so all the other model parameters are fixed
in the process. Since outflow and convection are taken into
account in our model, $\dot{M}$ is a function of radii. 
We therefore define $\eta \equiv \dot{M}_{\rm jet}/
\dot{M}(5R_s)$, where $\dot{M}(5R_s)$ is the accretion rate at 5  
Schwarzschild radii. When the luminosity is
relatively high, such as at the point ``A'' in Fig. 1,
$L_{\rm X}$ is dominated by the accretion flow (ref. Fig. 2 in YCN05). 
With the decrease of $\dot{M}$, however, the contribution of 
the jet to $L_{\rm X}$  becomes more and more important.
This is because X-ray emission from the accretion flow scales 
as $L_{\rm X,acc} \propto \dot{M}^q$ with $q \sim 2$ (see below for
details), while that from the jet is due to the optically-thin
synchrotron emission and thus $L_{\rm X,jet} \propto \dot{M}$ 
(e.g., Heinz 2004).
The radio luminosity is always dominated by the jet.

We find that our results are not sensitive to all the model 
parameters except $\delta$, which describes the fraction of
the viscously dissipated energy in directly heating electrons in the
hot accretion flow (YCN05). We first consider the case of $\delta=0.5$,
the value required in the detailed modeling of 
Sgr A* by a most updated ADAF model (Yuan, Quataert \& Narayan 2003).
The solid line in Fig. 2 shows the dependence of $\eta$ on $\dot{M}$.
We can see from the figure that
in this case to fit the radio---X-ray correlation, $\eta$ must be 
a strongly decreasing function of
$\dot{M}$.  If $\eta = const.$,
the predicted radio---X-ray correlation index would be $\xi_{\rm RX}
\sim 1.3-1.4 \gg 0.7$. This seems to be at odds with the result of MHD03.
We find that the discrepancy is mainly due to their adoption of
a smaller $\delta$ ($=0.3$). Following the notations in MHD03,
if $L_{\rm R}\propto \dot{M}^{\xi_{\dot{M}}}$ and $L_{\rm X,acc}
\propto \dot{M}^q$ ($L_{\rm X,acc}$ is the X-ray luminosity emitted from the
accretion flow), the correlation index $\xi_{\rm RX}=\xi_{\dot{M}}/q
= 1.4/q$ ($\xi_{\dot{M}}=1.4$: see Heinz \& Sunyaev 2003), 
if $\eta$ is assumed to be constant. 
MHD03 find that $q \sim 2.3$ for $\delta=0.3$.  
We do the calculations using $\delta=0.3$ and 
find that our result is in general agreement 
with MHD03. But for $\delta=0.5$, we find
$q\approx 1.1, 1.4$, and $1.8$ for $\dot{M}$ in the ranges of 
($5\times 10^{-2}, 2.5\times 10^{-3}), (2.5\times 10^{-3}, 5\times 10^{-4}$)
and $(5\times 10^{-4}, 1\times 10^{-4})$ in units of $L_{\rm Edd}$, 
respectively.
The reason for the difference in $q$ for different $\delta$
is as follows. The value of $L_{\rm X,acc}$ depends 
on the density $n_e$ and temperature $T_e$. 
With the decreasing of $\dot{M}$, the density decreases but $T_e$ increases. 
For larger $\delta$, $T_e$ increases faster thus $q$ is smaller. 
For comparison, we also calculate the case of $\delta=0.01$ and find 
that $\eta$ is nearly constant. In this case, 
$q\sim 2.4$, so the correlation index $\xi_{\rm RX}= 1.4/q \sim 0.6$
for a constant $\eta$, which is very close to 0.7.
Another reason for the discrepancy between our result and that of MHD03
is that we take into account the contributions of both the accretion flow
and jet (MHD03 attribute $L_{\rm X}$ only to
the accretion flow), which results in a smaller ``effective''
$q$. The third (minor) reason is that we consider the effects of
outflow and convection.

\subsection{The Steepening of Correlation and the Quiescent State of 
BHCs}

In the following we investigate
the correlation below the point ``B'' in Fig. 1. 
We assume that the jet persists and the physics of jets does not
change significantly at low luminosities.
We extrapolate the derived $\eta(\dot{M})$
(which is approximately a power-law) to lower $\dot{M}$ and calculate 
$L_{R}$ and $L_{X}$ for different $\dot{M}$.     
The segments ``B-C-D'' in Fig. 1 show our predicted radio---X-ray correlation
at low $\dot{M}$. It is interesting to see that 
below a certain luminosity, represented
by the point ``C'' in Fig. 1, the correlation deviate from the 
extrapolation of the observed radio---X-ray correlation
and approach asymptotically the segment ``DE'', 
\be 
\left(\frac{L_{\rm R}}{L_{\rm Edd}}\right) = 10^{-4.1}\left(\frac
{L_{\rm X}}{L_{\rm Edd}}\right)^{1.23}.
\ee
The segment ``DE'' shows the correlation of a pure-jet model, with the
radio/X-ray emission being due to the optically thick/thin synchrotron
emission of the electrons in the jet.
The normalization of the segment ``DE'' is determined by 
the results of modeling the outburst state of XTE J1118+480 (YCN05).
(The point ``D'' represents the emission from the jet in
XTE J1118+480 at the quiescent state, see Fig. 3(a)).
The change of the correlation is because, 
as we stated above, the contribution of jet to 
the total X-ray emission is becoming more and more important compared to
that of the accretion flow as $\dot{M}$ becomes smaller. 
Below a certain $\dot{M}$ (the point ``C'' in Fig. 1),
the X-ray luminosity 
will be completely dominated by the jet and thus the correlation 
of the system will follow that of the pure jet model.
This prediction is particularly relevant to the quiescent state of 
BHCs, because their X-ray luminosity is typically $\la 10^{-6}
L_{\rm Edd}$ (Kong et al. 2002; McClintock et al. 2003). 
We will discuss it further in \S 3.

The index of the correlation of a pure-jet model 
is $\sim 1.23$. This is in general agreement with Heinz (2004),
where he obtained $\xi_{\rm RX,jet}\approx 1.4$.
On the other hand, Markoff et al. (2003) obtained 
$\xi_{\rm RX,jet}\approx 0.7$. We find that the discrepancy is mainly
because Markoff et al. did not take into account the cooling break 
in the electrons energy distribution, as also pointed out by Heinz (2004).
\footnote{The effect of cooling break is correctly included in other 
jet models, e.g., Markoff, Falcke \& Fender 2001.}
We should note that our results are 
not very sensitive to the exact form of $\eta$. We examine the two
cases, $\delta=$ 0.5 and 0.01, as well as one
in which the value of $\eta$ at the lowest $\dot{M}$ in Fig. 2
is used. We find that the result remains qualitatively the same, 
although the exact location of ``C'' and the slope of the segment 
``BC'' in Fig. 1 are slightly different. 
So we conclude that the change in slope from ``AB'' to ``DE'' is robust.
The location of the intersection point ``C''
mainly depends on the normalization of the two segments. 
For different BHCs, the normalization of ``AB''
may vary by a factor of $\sim 5$ (Gallo, Fender \& 
Pooley 2003). The uncertainties in the normalization
of ``DE'' have two origins. One is from the jet model for a single source.
Unlike the accretion flow, the jet parameters are not well constrained
and there are some degeneracy. However, we find that because
of the excellent observational data of XTE J1118+480, the arisen uncertainty 
in the normalization is not large.
The uncertainty arisen from various sources  
depends on the diversity in the jet properties  
such as its velocity. These quantities are poorly constrained currently,
but we feel they should not differ too much among various sources.

\subsection{Extension from BHCs to AGNs}

While the radio---X-ray correlation index 
does not depends sensitively on the 
black hole mass $M$\footnote{From BHCs to AGNs, the synchrotron peak
from the ADAF will move from optical to radio. 
Depending on the value of $\eta$, the contribution of the ADAF to 
$L_{\rm R}(8.6 {\rm GHz})$ could become important when
$L_{\rm X}$ is very low. In this case, 
the correlation index will become smaller. At frequencies far below 
8.6 GHz, however, this effect is not important.}, as shown by
Heinz (2004), the normalization does.
To extend our result to AGNs, we need the dependence of the correlation
on the mass of the black hole. At relatively high luminosities (above
the point ``C'' in Fig. 1), MHD03 found, from analyzing
a sample of AGNs and BHCs,
\be
{\rm log} L_{\rm R}=0.6^{+0.11}_{-0.11}~{\rm log}
L_{\rm X}+0.78^{+0.11}_{-0.09}~{\rm log}\left({M}/{\msun}
\right)+7.33^{+4.05}_{-4.07}. 
\ee
Neglecting the uncertainties, we re-write this correlation as
\be
{\rm log} \left(\frac{L_{\rm R}}{L_{\rm Edd}}\right)=0.6~{\rm log}\left(
\frac{L_{\rm X}}{L_{\rm Edd}}\right)+0.38~{\rm log}\left(\frac{M}{\msun}
\right)-7.926.
\ee
Eq. (4) is almost identical to eq. (1), but with 
additional dependence on $M$ included.

Using a jet model, Heinz (2004) investigated the correlation 
index between $L_{\rm R}$ and $M$, $\xi_{\rm RM,jet}$, and found that
$\xi_{\rm RM,jet}\sim 0$. We use our pure jet model to  
calculate the value of $\xi_{\rm RM,jet}$.\footnote{Our 
jet model developed in YCN05 also works for large $M$, because the basic 
physics of jet should not depend on $M$, and the dependence of quantities (such 
as the frequency of the cooling break) on $M$ 
have been self-consistently taken into account in our jet code.}
We calculate the radio luminosity
$L_{\rm R}$ at various $M$, adjusting $\dot{M}_{\rm jet}$ to keep
$L_{\rm X}$ constant, and  obtaining the value of 
$\xi_{\rm RM,jet} = \partial {\rm log}L_{\rm R}
/\partial {\rm log}{M} \sim 0.25$. This result is similar to that obtained by 
Heinz. Therefore, eq. (2) can be generalized as,
\be 
{\rm log} \left(\frac{L_{\rm R}}{L_{\rm Edd}}\right)=1.23~{\rm log}\left(
\frac{L_{\rm X}}{L_{\rm Edd}}\right)+0.488~{\rm log}\left(\frac{M}{\msun}
\right)-4.53, 
\ee 
or equivalently,
\be
{\rm log} L_{\rm R}=1.23~{\rm log}
L_{\rm X}+0.25~{\rm log}\left(M/\msun
\right)-13.45.
\ee
This equation describes the segment ``DE'' in Fig. 1, with additional
dependence on $M$ included. From eqs. (4) and (5), we can estimate the
X-ray luminosity at the point ``C'' (which is very close to the intersection 
point in Fig. 1),
\be
{\rm log}\left(\frac{L_{\rm X,crit}}{L_{\rm Edd}}\right)=-5.356-
0.17~{\rm log}\left(\frac{M}{\msun}\right).
\ee

\section{Observational tests}

\subsection{Radio---X-ray correlation at low luminosities}


Jonker et al. (2004) obtained (nearly) simultaneous
radio and X-ray fluxes of XTE J1908+094 during the decaying phase of
an X-ray outburst. Their X-ray measurements were taken on 2003 
March 23, April 19, and May 13, but the radio measurements only on
March 25 and April 12. We fit the X-ray fluxes with a parabola and
estimate the X-ray flux for April 12 from the best-fit curve. Similarly,
we obtain the radio flux for March 23 by linearly interpolating the
March 25 and April 12 measurements. From the measured and estimated
radio and X-ray fluxes for march 23 and April 12, we derive a
radio---X-ray correlation index, $\xi_{\rm RX}\approx 1.28$, which is
significantly different from $\xi_{\rm RX}\approx 0.7$.\footnote{
If we estimate the X-ray flux on April 12 by linearly interpolating the
March 25 and April 19 (or April 19 and May 13) measurements, the correlation
index would be $\xi_{\rm RX}\approx 1.48$ (or 1.0). If we estimated the 
X-ray flux on April 12 by assuming an exponential decay in X-ray 
flux with time between March 23 and April 19, as Jonker et al. (2004) did,
the correlation index would be $\xi_{\rm RX}\approx 1.12$.}
Jonker et al. (2004) speculated that the discrepancy
may imply that, different from other BHCs, the accretion flow in this source
is in the form of a standard thin disk rather than an ADAF, 
even at low luminosities.

Given that our predicted value for the correlation
index is consistent with the range allowed by the J1908+094 data,
however, we believe that a more likely
scenario for the steeper radio--X-ray correlation is that the X-ray
emission of XTE J1908+094 is already dominated by the jet at the observed
X-ray fluxes. If our explanation is correct, the X-ray luminosity of
the source would have to be below the critical value (as indicated by
the point ``C'' in Fig. 1), i.e., $L_{\rm X} \la L_{\rm X,crit}\approx 10^{-5.5}
L_{\rm Edd}$. The mass of the compact object in XTE J1908+094 is not known.
Assuming a mass of $10\msun$, we found that its X-ray luminosity would be
$L_{\rm X}\sim 8\times 10^{-4}(d/8.5{\rm kpc})^2L_{\rm Edd}$ on March 23,
and $\sim 3\times 10^{-4}(d/8.5{\rm kpc})^2L_{\rm Edd}$ on April 12,
which implies that the source would have to be very nearby, $d \sim 1$ kpc.
It remains to be seen whether this is the case. At present, the distance
to the source is poorly constrained, as pointed out by Jonker et al.
(2004). We should stress that due to the uncertainty in the location 
of the point ``C'', the uncertainty in $d$ is significant.

As for AGN, the observations of M31 seem to provide evidence that
supports our predictions. The source was not 
included in the sample used by MHD03,
presumably because the X-ray data were not available at the time.
In this source, the mass of the black hole is $10^{7.5}\msun$. The
radio luminosity of the source (at 3.6 cm) is $ 10^{32.2}$
and $10^{32.37} \ergs$ based on two different observations (Crane et al.
1992; 1993). The X-ray luminosity is very weak, $L_{\rm X}
\sim 10^{35.5}\ergs \sim 10^{-3.5}L_{\rm X,crit}$
($\sim 2.5 \sigma$ detection; Garcia et al. 2004). 
So this source is very appropriate for testing our
prediction. From $L_{\rm X}$ and $M$, eq. (3) predicts that the radio 
luminosity is $\sim 10^{34.45}\ergs$,
which is $\sim 100$ times higher than the observed value, while eq. (6)
predicts a value of $\sim 10^{32.2}\ergs$, which is in good
agreement with the observation. The spectral fitting result is shown
in Fig. 3(b). In addition,  
Garcia et al. (2004) estimated the value of $\dot{M}$ to be 
$\dot{M}\sim 6 \times 10^{-6}\dot{M}_{\rm Edd}$. But the X-ray luminosity
predicted by an ADAF with such a $\dot{M}$ is only $\sim 10^{31}\ergs$,
which is about 4 orders of magnitude lower than the observed value,
as shown by the dashed line in Fig. 3(b). On the
other hand, we find that to produce the observed $L_{\rm X}$ by a jet,
the required $\dot{M}_{\rm jet}\sim 5\times 10^{-9}\dot{M}_{\rm Edd}
\ll 6 \times 10^{-6}\dot{M}_{\rm Edd}$, which is reasonable. 
Of course, the X-ray detection by Garcia et al. (2004) needs confirmation.
 
Most sources in the sample of MHD03 are observed at relatively high
X-ray luminosities which are not good for testing our predictions. 
Here we briefly summarize the results on the few sources
in MHD03 that satisfy $L_{\rm X} \la 0.1 L_{\rm X,crit}$. 
We should keep in mind that large uncertainties exist in the
normalizations of both correlations in eqs (3) and (6) for individual sources.

\noindent
{\em NGC 2841}. $M=10^{8.42}\msun$, $L_{\rm X}=10^{38.26} \ergs
\approx 0.03 L_{\rm X,crit}$, and
$L_{\rm R}=10^{36} \ergs$. Eq. (3) predicts $L_{\rm R}=10^{36.9}\ergs$,
nearly 10 times higher than observation,
while eq. (6) predicts $L_{\rm R}=10^{35.9}$, which is close to the observed
value.

\noindent
{\em NGC 3627}. $M=10^{7.26}\msun$, $L_{\rm X} < 10^{37.6}\ergs\approx 
0.07 L_{\rm X,crit}$, $L_{\rm R}=10^{36.74}\ergs$\footnote{We recalculate 
$L_{\rm R}$, using a new distance consistent with that used in 
calculating $L_{\rm X}$.}. Eqs. (3) and (6) predict 
$L_{\rm R} < 10^{35.55} \ergs$ and
$L_{\rm R} < 10^{34.76} \ergs$, respectively, both of which are significantly 
smaller than the observed value. 

\noindent
{\em Sgr A*}. $M=10^{6.41}\msun$, $L_{\rm X}= 10^{33.34}\ergs \approx 
10^{-4.8} L_{\rm x,crit}$, and $L_{\rm R}=10^{32.5}\ergs$. Like M31, it
should also be a good source to test our prediction, given its extremely
low $L_{\rm X}$. The predicted radio luminosity from eq. (3) 
($L_{\rm R}\sim 10^{32.3}\ergs$) is much closer to the observed value than that 
from eq. (6) ($L_{\rm R}\sim 10^{29.3}\ergs$), which is opposite to 
our expectation. On the other had, it is well known that
Sgr A* is a special radio source (e.g., Falcke \& Markoff 2000).
Unlike the typical core-jet AGNs, Sgr A* is observed to be 
quite compact (e.g., Lo et al. 1998). One possibility is that there 
is no jet in Sgr A* thus our assumption of the existence of jets
fails. In this case, the radio emission in Sgr A* may come from nonthermal 
electrons in the ADAF (Yuan, Quataert \& Narayan 2003). 
If a jet does exist in Sgr A*, the power-law energy distribution of electrons
in the jet ($N(\gamma)\propto \gamma^{-p}$) must be unusually steep, e.g., 
$p>3$\footnote{Both the theoretical studies to shock acceleration
and the observed optically-thin radio synchrotron spectra of extended 
suggest $p\ga 2$. We use $p\approx 2.2$ in the present paper, as in YCN05.}, 
as argued by Falcke \& Markoff. Such a steep distribution results in an 
unusually high radio/X-ray ratio, consistent with 
the observed low luminosities at infrared and X-ray bands. 
It may be instructive to compare Sgr A* to  M31. 
Compared to M31, the mass of the black hole in 
Sgr A* is 10 times lighter, but $L_{\rm R}$ 
is even higher, and $L_{\rm X}$ is more than 100 times lower.

\noindent
{\em M32}. $M=10^{6.4}\msun$, $L_{\rm X}=10^{35.97}\ergs\approx 10^{-2.1}
L_{\rm x,crit}$. This would be another good source to test our predictions, but
unfortunately there is only an upper limit on $L_{\rm R}$, 
$< 10^{33.3}\ergs$. Eq. (3) predicts $L_{\rm R}=10^{33.9}\ergs$, 
which seems a bit too high, while eq. (6) predicts $L_{\rm R}=10^{32.5}\ergs$,
which is in better agreement with the observed value. 
Future radio measurements may provide more stringent tests.

In summary, the current data from AGNs are so far inconclusive and more 
radio and X-ray observations to low-luminosity sources are required.

\subsection{Origin of X-ray emission in the ``quiescent state''}

We predict that below $L_{\rm x,crit}$, the X-ray spectrum 
should be dominated by the emission from the jet. This prediction 
provides  a good theoretical frame for understanding an otherwise 
puzzling observational result on M87.
The X-ray emission of M87 is usually
modeled by an ADAF (e.g., Fabian \& Rees 1995; Reynolds et al. 1996). However,
a subsequent {\em Chandra} observation strongly implies that the
emission is dominated by the jet, as argued by Wilson \& Yang (2002)
based on the similarity of the X-ray spectra between the nucleus and 
jet knots. The jet dominance in M87 is, in our model, 
because its X-ray luminosity
$L_{\rm X} \sim 0.8 L_{\rm X,crit}$.

Another way to test our prediction is therefore to examine the shape 
of the X-ray spectrum in the ``quiescent state'' (defined here 
as black hole sources with $L_{\rm X}\la L_{\rm X,crit}$). 
In general the X-ray spectrum of a jet emission is roughly of  
a power-law shape. On the other hand, if the emission is dominated by 
an ADAF, as proposed by Narayan, McClintock \& Yi (1996) for the 
quiescent state of BHCs, the X-ray spectrum should be curved 
due to the Compton scattering by thermal electrons when $\dot{M}$ is very low,
as shown, e.g., in McClintock et al. (2003) 
and in Fig. 3 (a) of the present paper for the quiescent state
of XTE J1118+480. Unfortunately, the X-ray data of black hole
sources in ``quiescent state''  are not of sufficient
quality to discriminate the models. Thus this important test awaits 
deep X-ray observations with state-of-the-art instruments like 
those on {\em XMM-Newton}. 

Fig. 3 (a) shows our prediction on the 
quiescent state spectra of XTE J1118+480. In the model,
the mass loss rate of the jet is $\dot{M}_{\rm jet}=6\times 10^{-8} 
\dot{M}_{\rm Edd}$, which is assumed to be
$\sim 15\%$ of the accretion rate in the underlying ADAF. 
Except for $\dot{M}$ and $\dot{M}_{\rm jet}$, 
all other model parameters remain the same as in YCN05.
We can see from the figure that 
the X-ray emission of the quiescent state is dominated by the jet.
We predict a power-law X-ray spectrum with photon index of $\sim 2$,
which is in good agreement with the current best fit of
the observational result (McClintock et al. 2003). We also note that the photon
indices of other quiescent BHCs are also $\sim 2$ (Kong et al. 2002;
McClintock et al 2003). Another issue we would like to mention,
as pointed out by McClintock et al. (2003), is the mass accretion rate
in the quiescent XTE J1118+480. Assuming that the 
optical emission comes from a truncated thin disk with an inner radius 
$R_{\rm tr}$, the value of $R_{\rm tr}$ can be determined 
from the optical flux, which is $\sim 1500 R_s$. 
Combining this result together with the disk instability 
theory for the outburst, we can estimate the  
mass accretion rate of the ADAF, which is 
$\dot{M}\la 10^{-6}\dot{M}_{\rm Edd}$. However, an
ADAF with such an accretion rate would under-predict
the X-ray flux by nearly four orders of magnitude (ref. Fig. 3 (a)).
On the other hand, if the X-ray flux is from the jet as we suggest above, 
there will be no such a problem, because this accretion rate 
is $\sim$ 20 times higher than the above $\dot{M}_{\rm jet}$.
Of course, if the optical flux is generated by the impact 
of the stream from the companion star on the disk surface, there will be no 
such a constraint on $\dot{M}$ (McClintock et al. 2003). 

\section{Discussion}

Fender, Gallo \& Jonker (2003; see also Gallo, Fender \& Pooley 2003)
compared the power of the jets, $P_{\rm jet}$ (as
inferred indirectly from the radio luminosity $L_{\rm R}$), and 
the X-ray luminosity $L_{\rm X}$ of BHCs. Extrapolating 
$L_{\rm R} \propto L_{\rm X}^{0.7}$ to low luminosities, they showed that
when the X-ray luminosity is below a critical value, 
$P_{\rm jet}$ should be greater than $L_{\rm X}$. 
The implication of this result is, however, not clear. 
For instance, it does not mean that the quiescent state X-ray emission
of BHCs is  dominated by the jets, which is what we conclude in the 
present work. Moreover, we predict that the radio---X-ray correlation
becomes much steeper at low luminosities. In addition, 
the outburst-state X-ray luminosity of 
XTE J1118+480 is far above their critical
luminosity ($L_{\rm crit} \sim 4\times 10^{-5}L_{\rm Edd}$),
so  $P_{\rm jet}$ should be much smaller than $L_{\rm X}$
according to their prediction. However, our calculation (YCN05)
shows that $P_{\rm jet} \approx 2 L_{\rm X}$, and Malzac, Merloni \&
Fabian (2004) obtain $P_{\rm jet} \approx 10 L_{\rm X}$.

In the quiescent state, BHCs seem to be much less luminous than 
their neutron star counterparts (e.g., Garcia et al. 2001;
McClintock et al. 2003). Narayan, Garcia \& McClintock (1997; see also 
McClintock, Narayan \& Rybicki 2004)
take this as evidence for the existence of event horizon in BHCs
for the following reasons. For neutron star systems, 
the energy stored in the accretion flow (ADAF) should eventually be released
as radiation upon impacting the solid surface of the neutron star.
The radiative efficiency is $\sim GM/R_*c^2\sim 0.15$. 
For BHCs, however, the energy stored in the ADAF simply
disappear into the event horizon of the black hole, so the
luminosity is expected to be much lower. Even if the luminosity of BHCs
is dominated by the emission from jets, this argument can still hold.
Turning the argument around, the systematic difference in the observed
X-ray luminosities of black hole and neutron star systems in the
quiescent state poses a constraint on our model.
In the jet-dominated case, the radiative efficiency of the whole system
will be $q_{\rm jet,rad}\eta$, with $q_{\rm jet,rad}$ being the 
radiative efficiency of the jet. To explain the difference of a 
factor of $\sim 100$ between the luminosities of BHCs 
and their neutron star counterpart (see Fig. 16 in McClintock et 
al. 2003), $q_{\rm jet,rad}\eta$ must be $q_{\rm jet,rad}\eta\sim
0.0015$. Given $q_{\rm jet,rad}\sim 0.05$ (YCN05)
and $\eta\sim 10\%$ or $1\%$ (see Fig. 2), we have $q_{\rm jet,rad}\eta
\sim 0.005$ or $0.0005$, which are comparable to the required value.

\section{Summary}

The main conclusions from this work can be summarized as follows:
1) Our accretion-jet model developed in YCN05 can re-produce the
observed radio---X-ray correlation with index of $0.7$ (Fig. 2).
2) Assuming that the jet persists, we predict that below a critical X-ray
luminosity ($L_{\rm X,crit}$) defined in eq. (7), the 
radio---X-ray correlation should turn steeper, from eq. (3) to eq. 
(6) (Fig. 1).
3) A related prediction is that the X-ray emission of a source is dominated
by that from the jet, when its X-ray luminosity is below
$L_{\rm X,crit}$. This is particular relevant to the X-ray emission of BHCs
in the quiescent state and some ``quiescent'' AGNs 
(whose X-ray luminosity $\la L_{\rm X,crit}$) (Fig. 3).

\begin{acknowledgements}
We thank Dr. J.E. McClintock for providing us with the data of XTE J1118+480.
This work was supported in part by NASA grant NAG5-9998.

\end{acknowledgements}



{} 

\clearpage

\begin{figure}
\epsscale{0.90}
\plotone{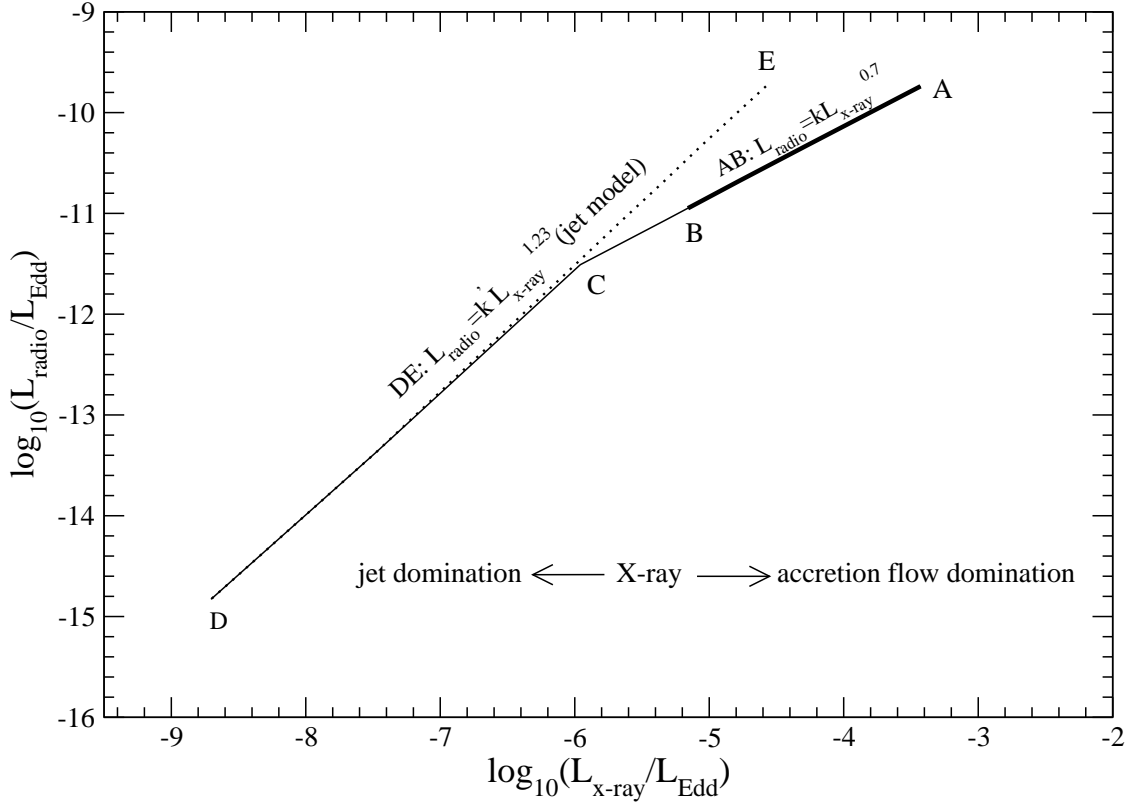}
\vspace{.0in}
\caption{The radio (8.6 GHz)---X-ray (2-11 keV)
correlation for BHCs. The observed correlation is
shown by the segment ``AB''. Segments
``BCD'' show the predicted correlation at lower luminosities, which 
approaches that of a pure-jet model, as shown by 
the segment ``DE''. Note that below point ``C'' 
($\sim 10^{-6}L_{\rm Edd}$), the X-ray emission is dominated
by the jet, and that the correlation steepens.}
\end{figure}

\begin{figure}
\epsscale{0.90}
\plotone{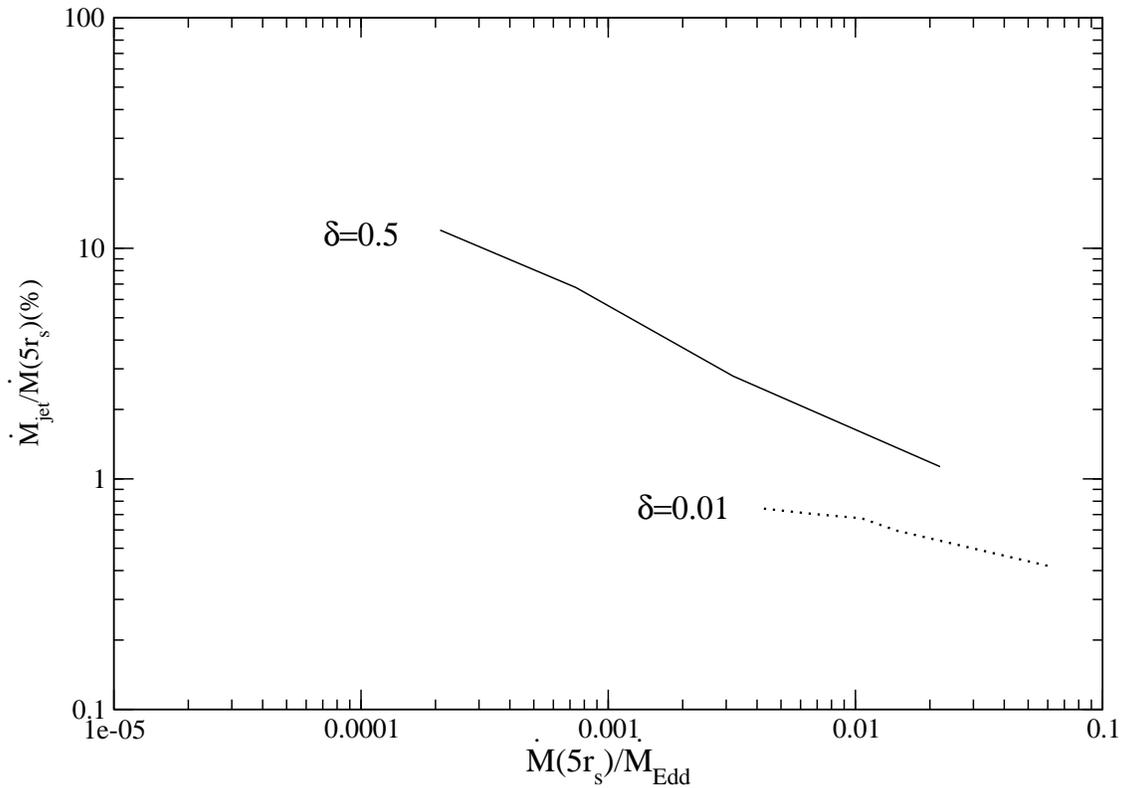}
\vspace{.3in}
\caption{The ratio of the mass loss rate in the jet ($\dot{M}_{\rm jet}$)
to the accretion rate of the ADAF at $\sim 5 r_s$ 
($\dot{M}(5r_s)$) as a function of the accretion rate.
The solid and dashed lines show results for two values of $\delta$
(the fraction of the viscously dissipated energy in directly 
heating electrons in an ADAF).}
\end{figure}

\begin{figure}
\epsscale{1.1}
\plotone{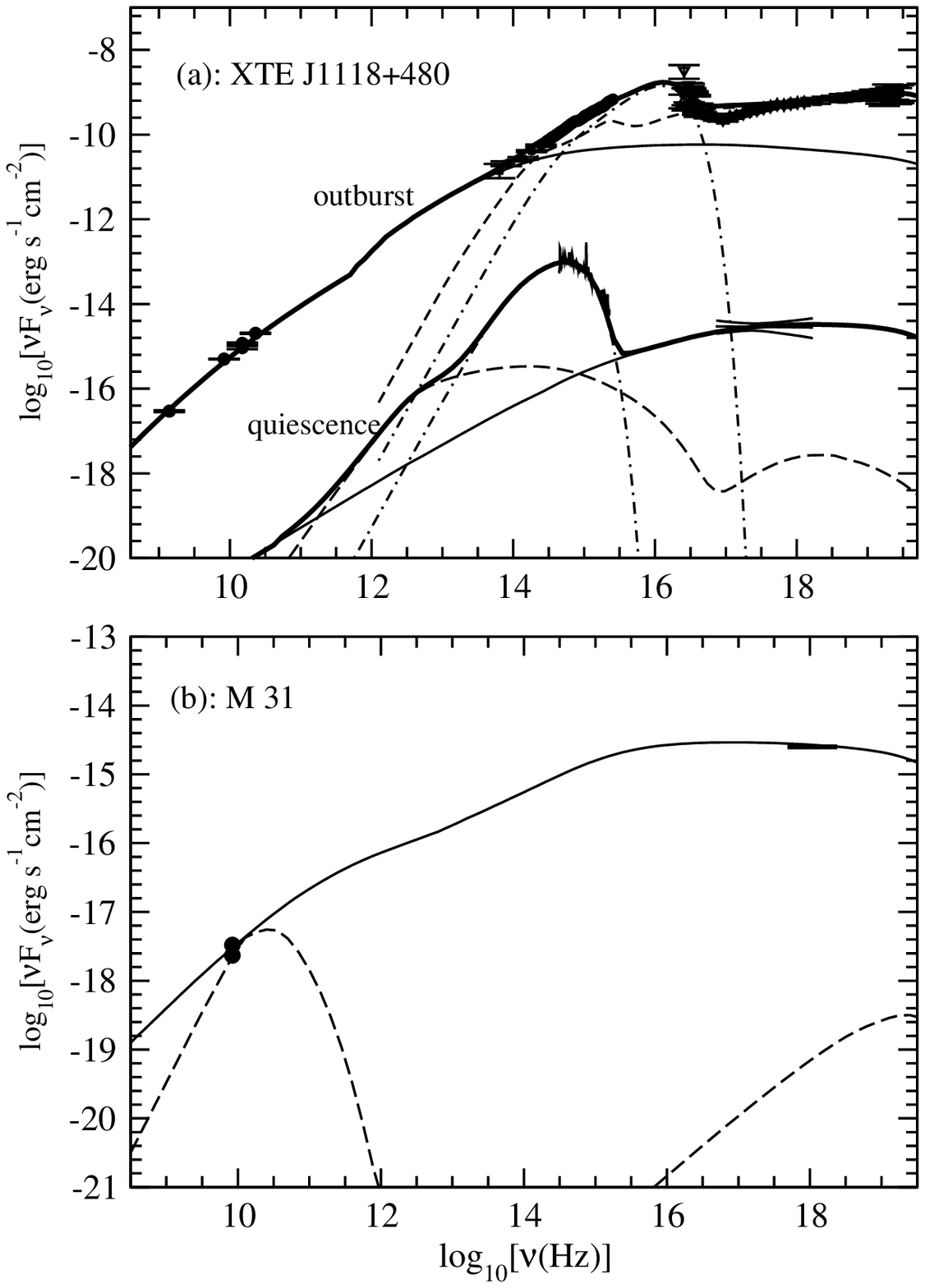}
\vspace{-.3in}
\caption{The accretion-jet model for two ``quiescent'' black hole sources.
(a) The quiescent state of XTE J1118+480.
The (optical and X-ray) data is from McClintock et al. (2003).
The thin solid line shows the emission of the jet, the dashed line for
the ADAF ($\delta=0.5$), and
the dot-dashed line that of a multi-temperature black
body component (e.g., a truncated disk; see McClintock et al. 2003 for details).
Their sum is shown by the thick solid line. The parameters are
$\dot{M}_{\rm jet}= 6\times 10^{-8}\dot{M}_{\rm Edd}$
and $\eta=15\%$.
Note that the X-ray emission is dominated by the jet. 
The model for the outburst state (YCN05)
is also presented for comparison purpose.
(b) ``Quiescent'' AGN---M31.
The radio data is from Crane et al. (1992; 1993) and the X-ray flux
from Garcia et al. (2004). A power-law X-ray spectrum with photon index of
$2$ is assumed. The solid line shows the emission 
of the jet while the dashed line for the ADAF ($\delta=0.01$). The 
parameters are $\dot{M}_{\rm jet}= 5\times 10^{-9}\dot{M}_{\rm Edd}$ 
and $\eta=1\%$. Again, the X-ray emission is dominated by the jet.}
\end{figure}


\begin{thebibliography}{}
\def\refpar{\hangindent=3em\hangafter=1}
\def\ref{\bibitem[]{587}}
\def\reference{\bibitem[]{588}}
\def\apj{ApJ}
\def\apjs{ApJS}
\def\mnras{MNRAS}
\def\aa{A\&A}
\def\aas{A\&A Suppl. Ser.}
\def\aj{AJ}
\def\araa{ARA\&A}
\def\nat{Nature}
\def\pasj{PASJ}
\def\pasp{PASP}

\bibitem []{} Blandford, R.D., K\"onigl, A. 1979, \apj, 232, 34

\bibitem []{} Corbel, S. et al. 2003, \aa, 400, 1007

\bibitem []{} Crane, P.C., Dickel, J.R., Cowan, J.J. 1992, \apj, 390, L9

\bibitem []{} Crane, P.C. Cowan, J.J.; Dickel, J.R., Roberts, D.A. 
1993, \apj, 417, L61

\bibitem []{} Esin, A. A., McClintock, J. E., \& Narayan, R. 1997,
ApJ, 489, 865

\bibitem []{} Fabian, A. C.; Rees, M. J. 1995, \mnras, 277, L55

\bibitem []{} Falcke, H., K\"ording, E., \& Markoff, S. 2004, \aa, 414, 895

\bibitem []{} Falcke, H., \& Markoff, S. 2000, \aa, 362, 113

\bibitem []{} Fender, R.P. 2004, To appear in 'Compact Stellar X-Ray Sources', 
eds. W.H.G. Lewin and M. van der Klis, Cambridge University 
Press (astro-ph/0303339)

\bibitem []{} Fender, R.P., Gallo, E. \& Jonker, P.G. 2003, \mnras, 343, L99
 
\bibitem []{} Gallo, E., Fender, R.P., \& Pooley G.G. 2003, \mnras, 344, 60

\bibitem []{} Gallo, E., Fender, R.P., \& Hynes, R.I. 2004, MNRAS, in press
(astro-ph/0410387)

\bibitem []{} Garcia, M.R., McClintock, J.E., Narayan, R. et al., 2001,
\apj, 553, L47

\bibitem []{} Garcia, M.R., Williams, B.F., Yuan, F., et al. 2004,
submitted to \apj (astro-ph/0412350)

\bibitem []{} Heinz, S. 2004, \mnras, in press (astro-ph/0409029)

\bibitem []{} Heinz, S., \& Sunyaev, R. A. 2003, \mnras, 343, L59

\bibitem []{} Jonker, P.G., Gallo, E., Dhawan, V., et al.
2004, \mnras, 351, 1359

\bibitem []{} Kong, A.K.H., McClintock, J.E., Garcia, M.R., 
et al. 2002, \apj, 570, 277

\bibitem []{} Lo, K.Y., Shen, Z.Q., Zhao, J.H., Ho, P.T.P.
1998, \apj, 508, L61

\bibitem []{} Malzac, J., Merloni, A., Fabian, A. 2004, \mnras, 351, 253


\bibitem []{} Markoff S., Falcke, H, \& Fender, R. 2001, \aa, 372, L25

\bibitem []{} Markoff S., Nowak, M., Corbel, S., Fender, R.,
Falcke, H. 2003, \aa, 397, 645

\bibitem []{} McClintock J.E., Narayan, R., Garcia, M. et al., 
2003, \apj, 593, 435

\bibitem []{} McClintock J.E., Narayan, R., \& Rybicki, G.B. 2004, 
\apj, 615, 402


\bibitem []{} Merloni A., Heinz, S., \& Di Matteo, T. 2003, \mnras, 345, 
1057 (MHD03)

\bibitem []{} Narayan, R., to appear in ``From X-ray Binaries to Quasars: 
Black Hole Accretion on All Mass Scales'', edited by T. Maccarone, 
R. Fender, L. Ho, to be published as a special edition of 
"Astrophysics and Space Science" by Kluwer (astro-ph/0411385)

\bibitem []{} Narayan, R., Garcia, M.R. \& McClintock, J.E. 1997, \apj, 478, L79

\bibitem[]{1339} Narayan, R., McClintock, J.E., Yi, I. 1996, \apj, 457, 821

\bibitem []{} Reynolds, C. S., Di Matteo, T., Fabian, A. C., et al. 1996, 
\mnras, 283, L111


\bibitem []{} Wilson, A. S.\& Yang, Y. 2002, \apj, 568, 133

\bibitem []{} Yuan, F., Cui, W., \& Narayan, R. 2005, \apj, 620, 905 (YCN05) 


\bibitem []{} Yuan, F., Quataert, E., Narayan, R. 2003, \apj, 598, 301

\end{thebibliography}
\end{document}